\documentclass[prb,preprint]{revtex4-1}

\usepackage[version=3]{mhchem} 
\usepackage{amsmath}
\usepackage{amsfonts}
\usepackage{graphicx}
\usepackage{hyperref}
\usepackage{color}
\newcommand{\relerr}[1]{\frac{{d #1}}{#1}}
\usepackage[normalem]{ulem}

\begin{document}

\title{Listen! It's a phase transition. \\ The sound of a shape-memory alloy.}

\author{Carlo Andrea Rozzi}
\affiliation{CNR --- Istituto Nanoscienze, via Campi 213/a, 41125 Modena, Italy}
\email{carloandrea.rozzi@nano.cnr.it}
\author{Annamaria Lisotti}
\affiliation{Dipartimento di Scienze Fisiche, Informatiche e Matematiche, Universit\`a di Modena e Reggio Emilia, via Campi 213/a, 41125 Modena, Italy}
\author{Guido Goldoni}
\affiliation{Dipartimento di Scienze Fisiche, Informatiche e Matematiche, Universit\`a di Modena e Reggio Emilia, via Campi 213/a, 41125 Modena, Italy}
\author{Valentina De Renzi}
\affiliation{Dipartimento di Scienze Fisiche, Informatiche e Matematiche, Universit\`a di Modena e Reggio Emilia, via Campi 213/a, 41125 Modena, Italy}
\affiliation{CNR --- Istituto Nanoscienze, via Campi 213/a, 41125 Modena, Italy}
\email{vderenzi@unimore.it}

\begin{abstract}
Shape-memory alloys exhibit a solid-to-solid phase transition that involves a temperature-driven rearrangement of their crystal structure and is responsible for their remarkable properties and numerous technological applications. Here, we propose a simple experiment that analyzes the sound emitted by a Ni$_{40}$Ti$_{50}$Cu$_{10}$ bar at different temperatures as it undergoes a transition between its austenite and martensite phases. We show that the phase transition, which occurs slightly above room temperature, can be qualitatively detected by the ear and quantitatively described using a very simple experimental setup and sound analysis tools. Such a sound-based investigation provides an unusual and engaging way to experimentally introduce solid-to-solid phase transitions, that is suitable for undergraduate courses.
\end{abstract}

\maketitle

\setlength{\parskip}{0pt} 

\section{Introduction}\label{sec:Intro}

Phase transitions (PTs) are characterized by a system's collective 
response to a small change of some external parameters, such as temperature, pressure, magnetic field or strain.
The familiar solid/liquid and liquid/gas transformations are everyday examples to conceptualize PTs that are often discussed in courses.
%
%
%
However, providing additional examples of PTs that can be experimentally investigated in an educational lab is valuable to illustrate the generality and ubiquity of these phenomena.

The martensitic PT represents such a pedagogical example for which we have designed several didactical activities \cite{nanolab}; it is a first order solid-to-solid PT occurring in so-called Shape Memory Alloys (SMAs) such as nickel-titanium alloys (nitinol) and its derivatives.\cite{NiTiCrystal, Def2} As shown in Fig.~\ref{fig:phases}, it consists in a change of the microscopic crystalline structure from cubic (austenite) to monoclinic (martensite), driven by temperature. 
The experimental study of the martensitic PT has several practical and pedagogical advantages. Nitinol is a solid, safe, easy to manage, inexpensive metal, that is available off-the-shelf. The transition occurs within a lab-suitable  temperature range (around $50^\circ$C). Although it will not be the subject of this paper, the transition underlies the astonishing shape-memory effect (SME), which is the ability of a deformed nitinol sample to return to its original forged shape when heated.\cite{video-SME} The SME is a quite popular topic for classroom demonstrations and science exhibits,\cite{Edu2, Edu3} with numerous technological applications~\cite{ReviewNiTiApplications} which may spark students' curiosity. 

Apart from the SME, the martensitic PT is hardly visible to the eye. However it has an important -- though less explored -- impact on the elastic and acoustic properties of the material. This can be readily observed by dropping a small nitinol bar on the floor, first in its austenite phase - i.e at temperatures above the transition - and then in its martensite phase (at temperatures below the transition).\cite{recordings} The austenite rings like a metallic object, characterized by a distinct pitch, while the martensite phase produces a muffled “thud” with no definite pitch, resembling that of a softer, non-metallic object.~\cite{storica} 

While it is a common experience for musicians to find their instrument slightly out of tune when they undergo environmental changes, in our case, a moderate temperature change induces a very large frequency shift, and turns a pitched idiophone (the nitinol bar in the austenite phase) into an unpitched one (in the martensite phase). 
This simple yet astonishing observation naturally raises the question of whether sonic data could be used to investigate the martensitic PT. Indeed, a few authors pursued this idea in a pedagogical context, but only qualitatively.~\cite{NiTiTemp, spectroscopy}

In this paper, we show that the sound produced when one hits a nitinol sample can be continuously monitored through the PT using an elementary experimental setup,
allowing quantitative correlation of the pitch and timbre of the perceived sound with the phase state, thus exposing the occurrence of the PT, though no visible macroscopic change in the material occurs. This illustrates how the material’s macroscopic properties originate at the nanoscale.
From an educational point of view, it is also important to note that this experiment draws a bridge between two usually distinct fields: acoustics and material science, which can be used to introduce damping, vibration and noise control in dynamical structures.

\section{Theoretical Background}

\subsection{Temperature dependence of the fundamental frequency in a thin metal bar}\label{sec:acoustic}

This section briefly discusses how temperature affects the fundamental frequency emitted by a bar that is not undergoing a phase transition. This frequency depends on the bar's dimensions (length $L$ and thickness $h$) and on the Young's modulus $E$ of the material (see details in Appendix A). If the motion is flexural (i.e. only along the direction parallel to $h$), then the fundamental frequency is
\begin{equation}
  \nu_0(T)\propto h^2(T)\sqrt{\frac{E(T)}{L^3(T)}},
\end{equation}
and therefore the fractional frequency change with temperature is
\begin{equation}
  \relerr{\nu} = 2\relerr{h}-\frac{3}{2}\relerr{L}+\frac{1}{2}\relerr{E}.
\end{equation}
At ordinary temperatures these changes are nearly linear with temperature, so the relationship simplifies
\begin{equation}\label{eq:deltanu}
  \relerr{\nu}=\frac{\alpha+\beta}{2}dT,
 \end{equation}
where 
$\alpha = (1/L)\ dL/dT = (1/h)\ dh/dT$ is the coefficient of the linear thermal expansion and $\beta = (1/E)\ dE/dT$ accounts for the Young's modulus first-order variation with temperature. For ordinary metals, the values of $\alpha$ and $\beta$ are quite small with $\alpha$ positive and $\beta$ negative, resulting in a change in the fundamental frequency that are very unlikely to be detected by ear when the bar cools gradually from 70$^\circ$C to 25$^\circ$C.

\subsection{The martensitic phase transformation in nitinol}\label{sec:PTnitinol}

Solid-to-solid PTs occur between two states of matter that have different crystallographic structures. Though these transformations consist of tiny atomic displacements -- if compared to the interatomic distances -- these subtle movements are responsible for dramatic changes in the macroscopic properties of the material. The martensitic PT is therefore a prototypical example of diffusiveless crystallographic transformations that are also called "military" transformations because they occur as ordered cooperative shear-like movements of atomic layers rigidly shifting (sliding) with respect to neighboring planes.

\begin{figure}[ht]
  \centering
  \includegraphics[width=\textwidth]{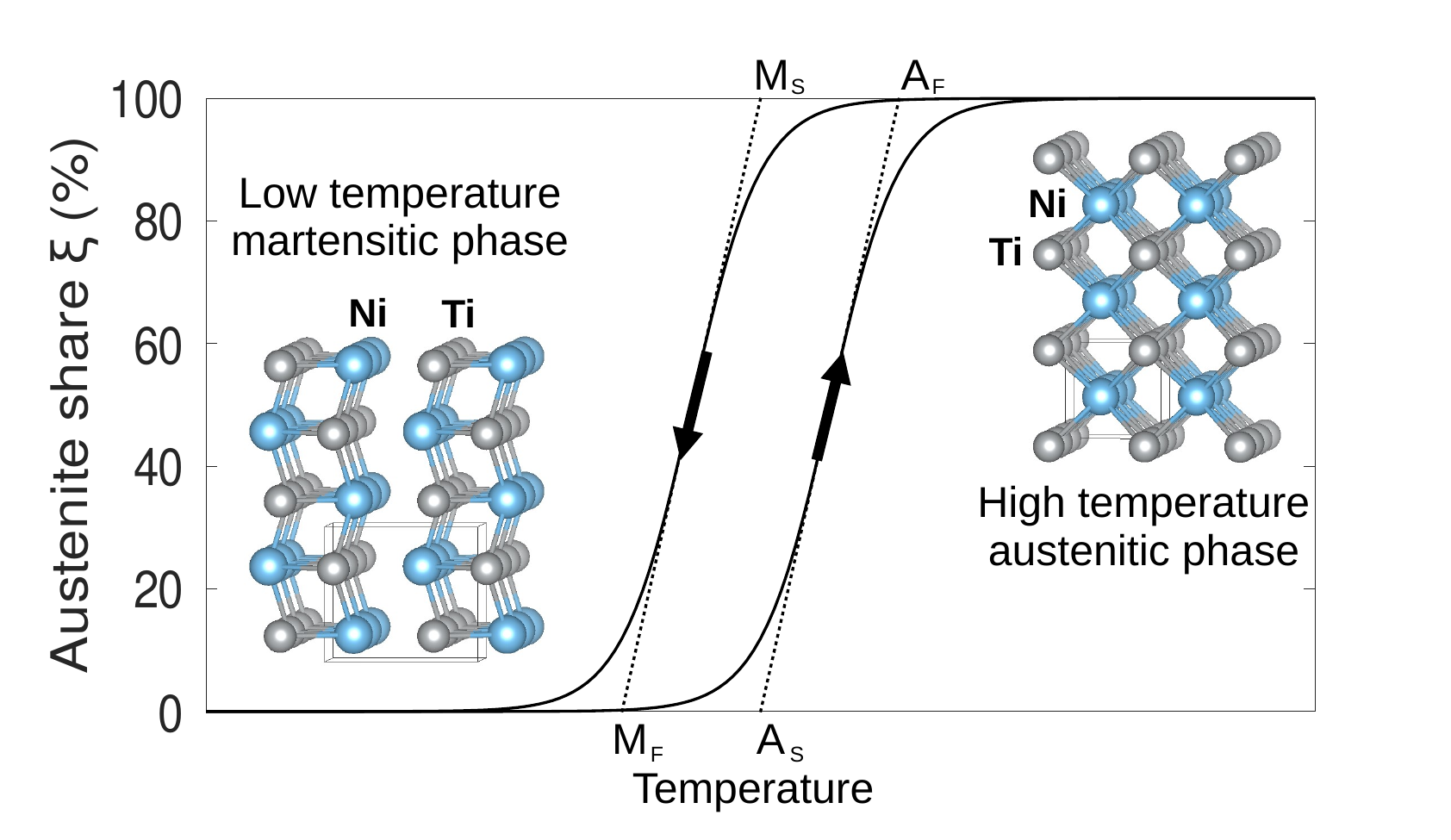}
  \caption{Theoretical austenite volume fraction throughout the phase transition. $M_S$, $M_F$ indicate the start and finish temperatures for the phase transition upon cooling; $A_S$, $A_F$ are the corresponding temperatures upon heating. Left inset: lattice geometry of the twinned martensite B19' phase (monoclinic crystal lattice, space group 11). Right inset: austenite B2 phase (cubic lattice, space group 221). The unit cells are delimited by the thin lines. The sticks are added to highlight the stacked patterns. They do not indicate chemical bonds.}
  \label{fig:phases}
\end{figure}

Although in this paper we deal with the ternary alloy Ni$_{40}$Ti$_{50}$Cu$_{10}$ (NiTiCu in the following), in Fig.~\ref{fig:phases} we present the microscopic structure of the two states involved in the martensitic PT for the simpler case of the NiTi alloy, where Ni and Ti atoms alternate, forming two interpenetrating lattices. At high temperature, the stable  crystal structure is a cubic austenite [Fig.~\ref{fig:phases}b)]. At low temperature the stable phase is the less symmetric monoclinic martensite [Fig.~\ref{fig:phases}a)]\cite{Note3}. The monoclinic symmetry of martensite means that it has three unequal crystallographic axes, and two of these axes are perpendicular to the third axis, but form an angle other than 90$^\circ$ to each other. This transition results in twinned domains with different orientations. 
Due to the hysteretic mobility of these variants, the martensite phase has a high damping capacity.\cite{Damping_capacity} Moreover, in the martensite phase, if stress is isothermally applied, the twinned domains may easily flip and the sample can be plastically deformed to any shape. The rearrangement between the twinned domains of the martensite and the austenite phases is at the origin of the SME occurring in nitinol and some other inter-metallic compounds~\cite{FirstSME,SMEAlloys,MostSME}. SME in itself is a fascinating phenomenon, but it is not of direct concern of the present investigation. Hence, we refer the interested reader to Refs.~\onlinecite{FirstSME,SMEAlloys} for a thorough discussion of this effect. Simulations of the microscopic mechanisms underlying the austenite/martensite transformation for the NiTi binary material can also be found in Ref.~\onlinecite{Ko2015}.

It is important for the following to highlight that -- unlike the better-known solid-liquid or liquid-gas transformations -- the temperature-induced austenite/martensite transformation is not isothermal;\cite{SMABook,PhaseTransfChap4} upon cooling, the austenite starts transforming into martensite at a ``martensite start" temperature $M_S$ and the process is completed at a ``martensite finish" lower temperature $M_F$ (see Fig.~\ref{fig:phases}). This can be explained taking into account the influence of internal stress on the transition temperature, in analogy with the role of pressure on the liquid-vapor transition temperature of water. Moreover, the presence of hysteresis - fingerprinting a first-order phase transition - can also be observed: upon heating, austenite starts forming at a "austenite start" temperature $A_S$, different from $M_F$, and the process is completed at a "austenite finish" temperature $A_F$, different from $M_S$. These temperatures are somewhat sample-dependent, but normally fall in the range of a few tens of Celsius degrees for NiTi compounds.\cite{NiTiTemp}

\section{Materials and methods}\label{sec:Methods}

We employed custom made bars donated by CNR-IENI~\cite{IENIbars} with nominal stoichiometry Ni$_{40}$Ti$_{50}$Cu$_{10}$. Similar nitinol sample bars are commercially available for a few tens of USD. 
 Our bar was $(20.0 \pm 0.1)$\,cm long, with an approximate rectangular cross section $(0.70 \pm 0.05)$\,cm $\times$ $(0.60 \pm 0.05)$\,cm, and a mass $(51 \pm 1)$\,g. At room temperature the sample was stable in its martensite phase. An additional cylindrical iron bar, $(20.2 \pm 0.1)$\,cm long and with a diameter of $(0.79 \pm 0.02)$\,cm and a mass $(77 \pm 1)$\,g, was used as a control sample.

\begin{figure}[ht]
  \centering
  \includegraphics[width=\textwidth]{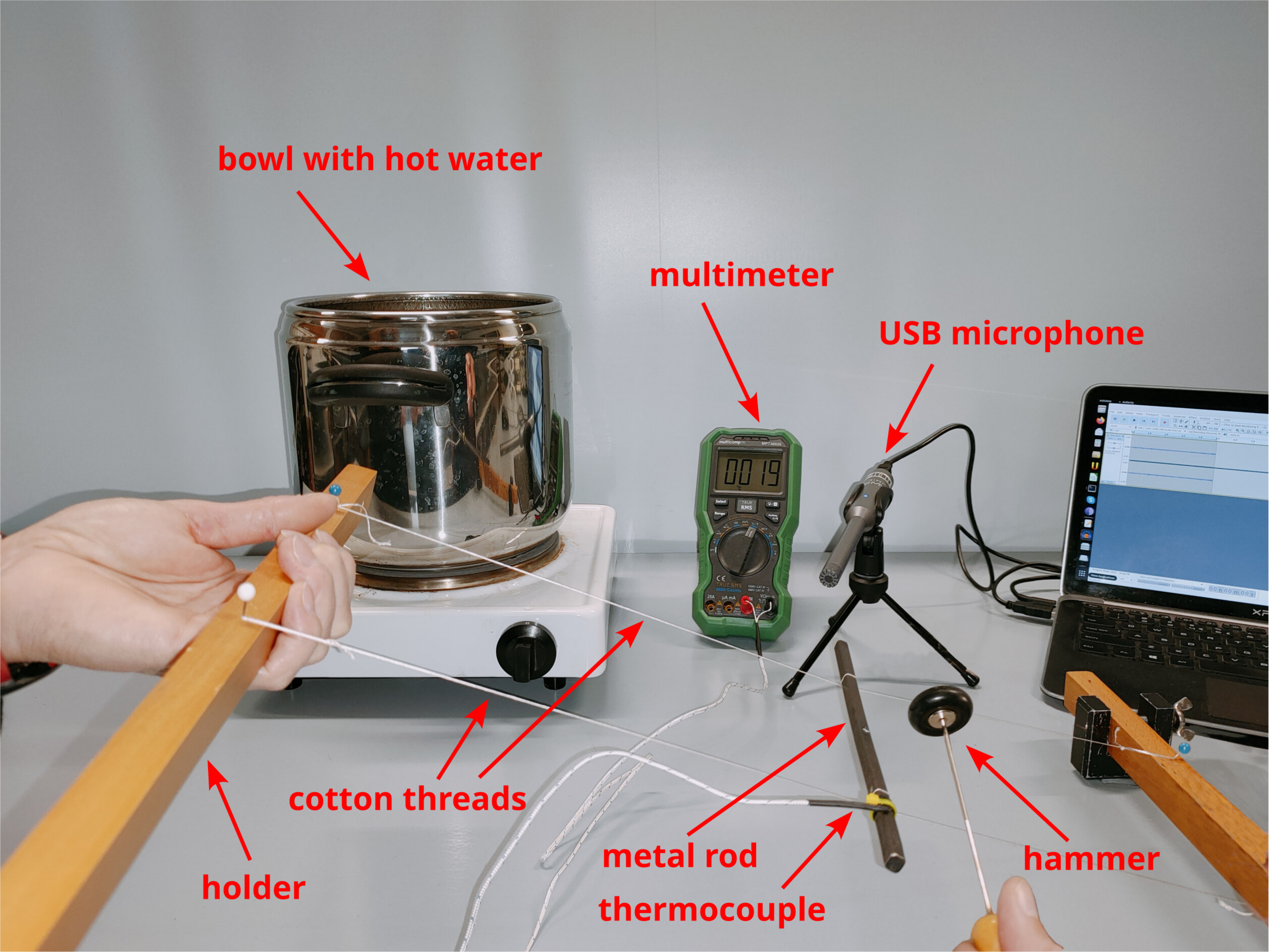}
  \caption{Photo of the experimental setup. The bar (either the nitinol or the iron bar) is tied to two rigid holders with cotton threads passing at the nodal planes of the bar. A thermocouple is fixed to the bar and the temperature is read by a multimeter. Using the holders, the bar is first immersed in a bowl containing water at a fixed temperature. Then, it is quickly extracted from water and held suspended with the holders. As the bar gradually cools, the operator may repeatedly take note of (or read aloud for recording) the temperature from the multimeter and gently hit the bar with a rubber hammer. The sound is recorded through a microphone connected to a computer to be analyzed.}
  \label{fig:setup}
\end{figure}

The experimental procedure is sketched in Fig.~\ref{fig:setup}. The bar (either the nitinol or the iron one) was hung from two cotton threads, allowing it to be submerged in hot water and then removed. The two cotton threads were tied to the bars at positions corresponding to the nodes of the first mode of free vibration (see App.~\ref{sec:Appendix}). The bars were immersed in water at $70^\circ$C before being quickly extracted and allowed to cool down to room temperature. While cooling, the bars were quickly and repeatedly hit with a rubber hammer in the middle, perpendicularly to their length, to excite flexural vibrations. A thermocouple was attached to each bar near one node to monitor its temperature.

The sound emitted by the bar was recorded with a USB microphone directly connected to a computer's sound card. The sound spectrum and the frequency of the fundamental mode were measured using the open-source software Audacity~\cite{Audacity}. The decay time of the hits was estimated via a linear fit of the log plot of the sound pressure trace (see App.~\ref{sec:Appendix2} for details). Alternative experimental setups are discussed in Sec.~\ref{sec:Experiment}.

\section{Results and Discussion}

\subsection{Fundamental frequency of NiTiCu and iron}\label{result_freq}

\begin{figure}[ht]
  \centering
  \includegraphics[width=\textwidth]{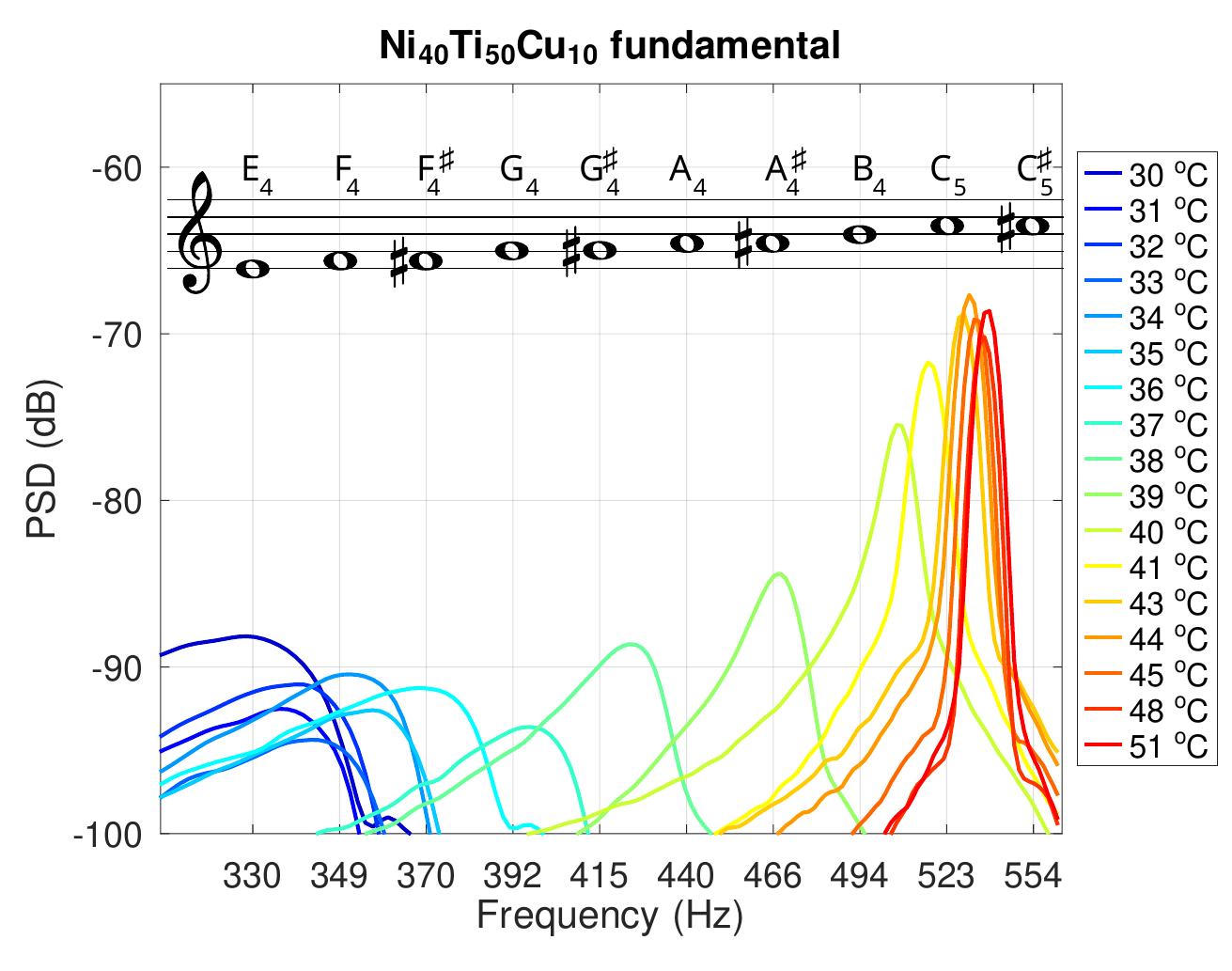}
  \caption{(Color online) Power spectral densities (PSD) of the fundamental mode of the NiTiCu bar at different temperatures. At the top we report the notes corresponding to the frequencies indicated by the vertical grid lines.}
  \label{fig:scale}
\end{figure}

The changes in the acoustic properties of the NiTiCu bar were recorded upon decreasing its temperature from 65$^\circ\mathrm{C}$ to 24$^\circ\mathrm{C}$, across the martensitic PT, and compared with those of the control iron bar in the same temperature range. While the iron bar sound remained about the same (we employed a bar with a fundamental frequency at about 876~Hz), the fundamental frequency of the NiTiCu bar decreased significantly as its temperature decreased (the recordings can be downloaded from the supplementary material). In Fig.~\ref{fig:scale} the recorded NiTiCu spectra are reported for selected temperature values, along with their corresponding scale, in musical notation, showing how the fundamental frequency decreased by almost 40$\%$ from roughly a sharp C$_5$ at 51$^\circ\mathrm{C}$ to a sharp E$_4$ at 30$^\circ\mathrm{C}$. For temperatures outside this interval the frequency remained quite stable. Moreover, for NiTiCu, other important changes occurred as the temperature decreased, namely: (i) the ringing time shortened, and (ii) the pitch became less sharply defined.

\subsection{Temperature dependence of elastic properties}\label{sec:Elastic}

\begin{figure}[ht]
  \centering
  \includegraphics[width=\textwidth]{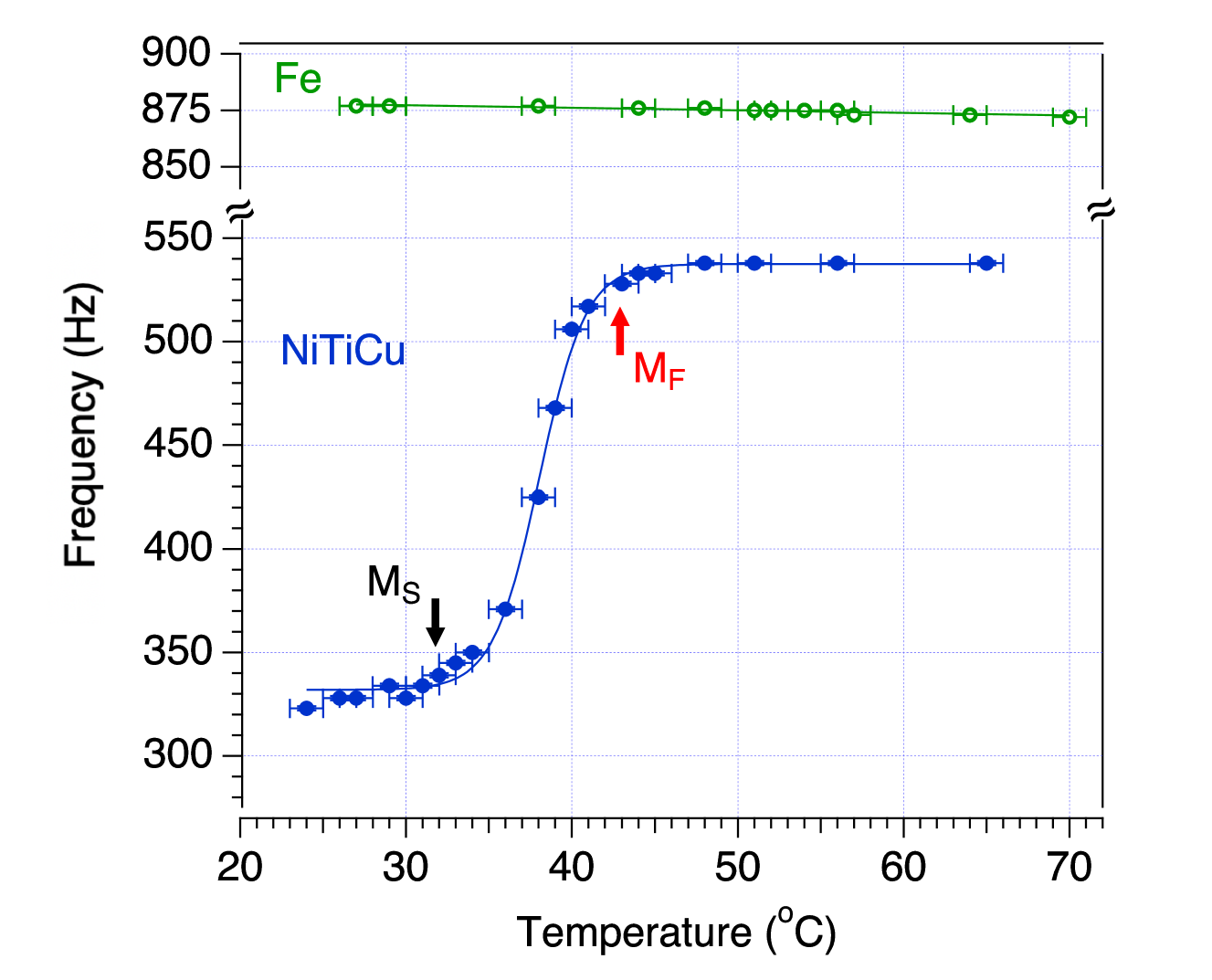}
  \caption{Temperature dependence of the fundamental vibration mode frequency of iron (top) and nitinol (bottom) rods upon cooling. The arrows mark the estimated Martensite start ($M_S$) and finish ($M_F$) temperatures at 43 and 32$^\circ$C respectively. The superimposed  lines are guides for the eye.}
  \label{fig:freq}
\end{figure}

The lowest peaks in the frequency spectrum of the iron and NiTiCu bars are reported in Fig.~\ref{fig:freq} at different temperatures. The ring tone of the iron bar was 871~Hz at 65$^\circ$C, and increased by 5~Hz as the temperature decreased to 20$^\circ$C. This variation was not detectable by ear, since most people can detect a difference in frequency of about 1\% when pure tones are played in succession. The slight decrease of the frequency with temperature, with a slope of -0.1~Hz.$^\circ$C$^{-1}$, is well described by Eq.~\ref{eq:deltanu}, using the values $\alpha = 10^{-5}\,^\circ\mathrm{C}^{-1}$ and $\beta = 10^{-4}\,^\circ\mathrm{C}^{-1}$ from Refs.~\onlinecite{LinThermExp, YoungTemp}.

The case of NiTiCu is qualitatively very different: The NiTiCu fundamental frequency dropped abruptly upon cooling between $M_S = 43^\circ$C and $M_F = 32^\circ$C. The corresponding average slope is 23~Hz.$^\circ$C$^{-1}$, more than two orders of magnitude larger than in the case of iron, and with a very unusual negative sign~\cite{normalExpansion}. This change is the fingerprint of the martensitic PT.

This sudden decrease in frequency upon cooling could be attributed to an abrupt change in the dimensions of the bar (and therefore of its density), the Young's modulus, or both. However, since the densities of martensite and austenite NiTiCu alloys differ by less than $5\%$,\cite{NoteUnitCell} we can safely conclude that density increase has a minor effect, and that the pitch drop should be attributed to a strong variation in the Young's modulus across the PT: austenite is much stiffer than martensite. 
 
\begin{figure}[ht]
  \centering
  \includegraphics[width=\textwidth]{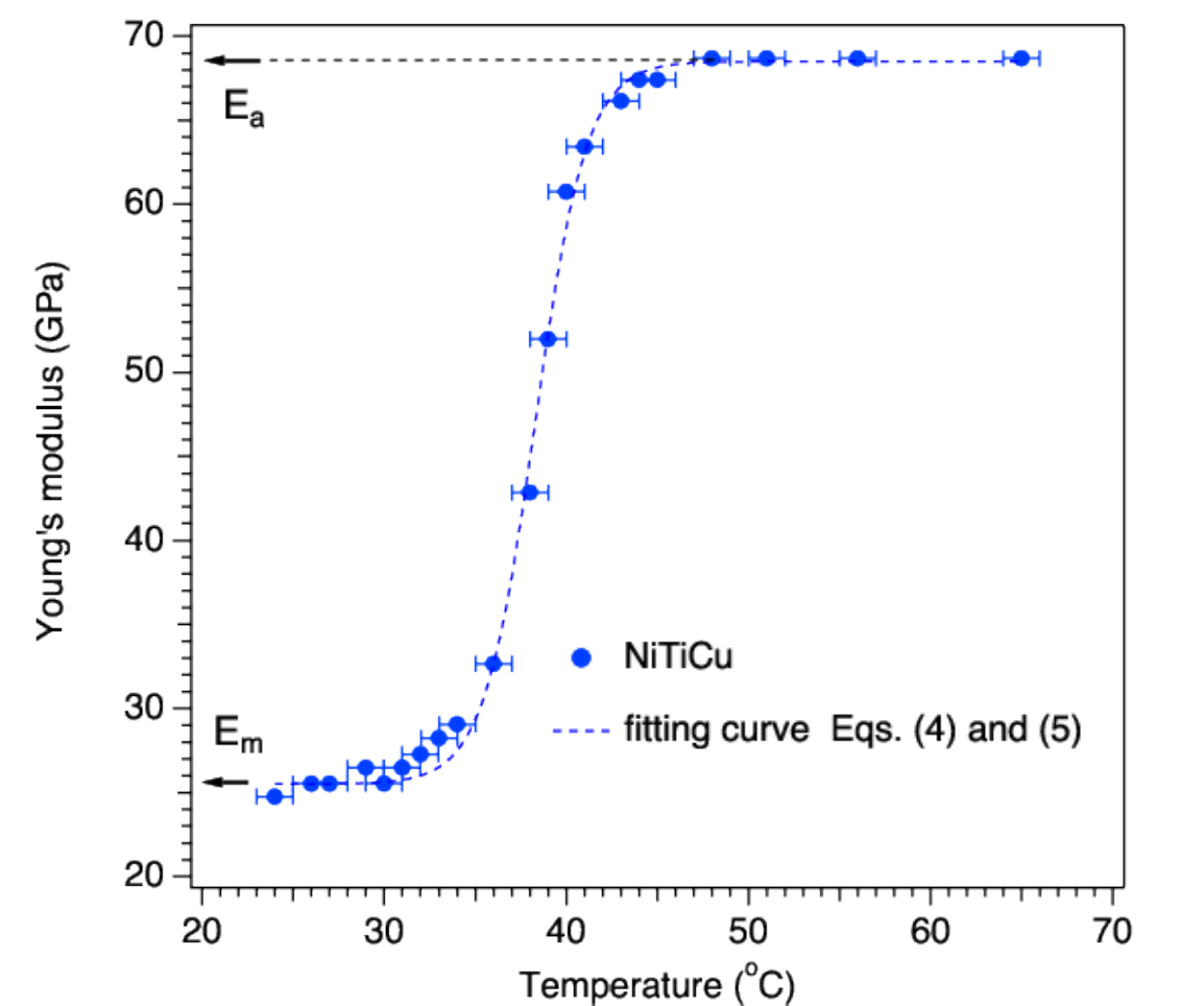}
  \caption{ Variation of Young's modulus as a function of temperature. The fitting curve (dashed line) is the logistic function, as indicated in eqs.~\ref{eq:Eeff} and \ref{eq:logistic} with parameters $T_m = 38.3^\circ$C and $k = -1.4^\circ$C$^{-1}$. The values of $E_a=(69\pm 1)$~GPa and $E_m=(26 \pm 1)$~GPa, obtained by the fitting procedure, are also indicated by horizontal arrows.}
  \label{fig:young}
\end{figure}

By using Eqs.~\eqref{eq:barfreq} and \eqref{eq:speed} of App.~\ref{sec:Appendix}, assuming constant density and geometrical parameters (measured for the martensite phase), we can estimate the Young's modulus, as shown in Fig.\ref{fig:young}. From the lowest and highest frequencies (respectively found at temperature lower than $M_F$ and higher than $M_S$) we obtain $E_m=(26\pm 1)$~GPa and $E_a=(69\pm 1)$~ GPa, in agreement with literature, with values from the literature of less than $40$~GPa for martensite and $60-90$~GPa for austenite in nitinol.\cite{E_niti, Ref_E}
In between these two extreme values, the temperature dependence of Young's modulus can be explained by recalling that the martensite and austenite phases coexist throughout the transition. For temperatures $M_S>T>M_F$, the upper bound for the effective Young's modulus can be interpolated between the martensite ($E_m$) and austenite ($E_a$) extreme values as
\begin{equation}\label{eq:Eeff}
  E_{\mathrm{eff}}(T) = \xi(T)E_a+ [1-\xi(T)]E_m = E_m + [E_a-E_m]\xi(T),
\end{equation}
where $\xi(T)$ is the fraction of austenite phase at temperature $T$, which is usually conveniently fit~\cite{Zotov2014} by the logistic curve
\begin{equation}\label{eq:logistic}
  \xi(T) = \frac{1}{1 + e^{k\left(T-T_m\right)}},
\end{equation}
with $T_m$ and $k$ free parameters representing the temperature midpoint and  the logistic growth rate, respectively. As shown in Fig.\ref{fig:young}, the experimental data can be nicely fitted using Eqs. \eqref{eq:Eeff} and \eqref{eq:logistic} with $T_m = 38.3^\circ$C and $k = -1.4^\circ\mathrm{C}^{-1}$.

At a microscopic level, the frequency shift can be understood in terms of rearrangements in the unit cell at the PT. The change in crystallographic structure leads to substantially different effective elastic potentials between atoms, i.e. different amount of elastic energy stored per unit volume in the material. In the austenite phase, deformation primarily involves direct bond stretching, and the cubic unit cell symmetry ensures that the atomic bonds resist deformation uniformly in all crystallographic directions. On the other hand, the less symmetric (monoclinic) martensite unit cell yields a shallower energy landscape, that allows for multiple variants (twinned or detwinned) to be energetically close to each other~\cite{YoungPhases}. These variants can reorient relatively easily under applied stress and absorb part of the strain both elastically (yielding a lower Young's modulus) and inelastically (yielding higher energy dissipation).

\subsection{The transition from unpitched to pitched sound}\label{sec:Pitch}

\begin{figure}[ht]
  \centering
  \includegraphics[width=0.85\textwidth]{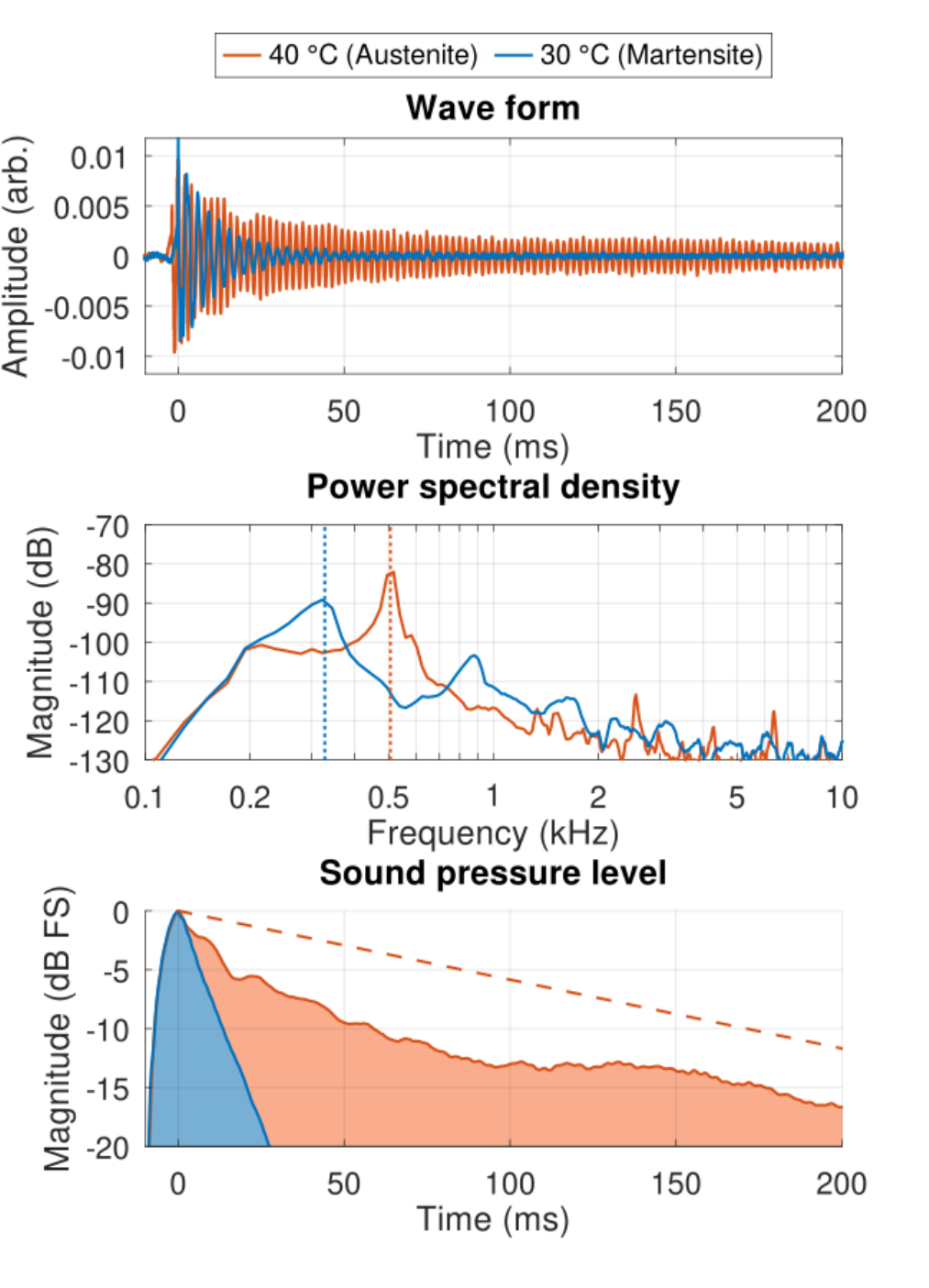}
  \caption{Two selected sounds of the NiTiCu bar at 40$^\circ$C (austenite) and 30$^\circ$C (martensite), i.e. close to the beginning and end of the martensitic PT. Top: time-domain wave forms. Middle: Power spectral densities highlighting the presence of a fundamental peak (marked with broken lines) and a few overtones. Bottom: normalized sound pressure level. The dB scale is relative to full scale (FS) values. The dashed line is the linear fit of the decay in the austenite phase (shifted to intercept the 0 dB magnitude).}
  \label{fig:damp}
\end{figure}

So far we have focused on the shift of the fundamental frequency emitted by the bar across the transition. However, when the NiTiCu bar is cooled, the fundamental frequency decreases, but the pitch also becomes less defined. Actually, below, typically, $M_F$, the sound emitted by the sample does not have a definite pitch. This change is audible and is reflected by the broadening of the pulse in the spectrum of the emitted sound, although regular oscillations are still observable in the recorded waveform (see Fig.~\ref{fig:scale} and ~\ref{fig:damp} top and middle panels).

The timbral change from a ``ring" sound into a ``thud" is correlated with the sound duration; a definite pitch is perceived only if the bar vibrates for a sufficient duration. This feature is shown in Fig.~\ref{fig:damp} (bottom panel), which depicts the envelopes of two recorded waveforms corresponding to the austenite and the martensite. The logarithmic scale in the vertical axis shows that the sounds decay at an exponential rate, with very different time constants for the two phases; the austenite decay time in this case is more than one order of magnitude longer than the martensite one (about 150~ms vs 10~ms).

The difference in decay time between the two phases is due to a drastic change in the damping coefficient of the bar's elastic oscillations. In general, damping coefficients result from mechanical energy dissipation, which can originate from various sources collectively referred to as "friction". Here, only internal friction is at play. External sources of friction, such as the viscous drag by air or friction at the support points, are the same in the martensite and austenite phases, as the transition does not significantly affect the macroscopic shape of the bar.~\cite{NotePorosity}

As the bar is hit and the pressure wave propagates through it, it undergoes a periodic sequence of stress and strain. Due to the material's hysteretic response to these constraints, elastic energy is converted into heat during the oscillations. Internal friction in metals and alloys is particularly sensitive to defects in the lattice and to their mobility. This picture fits the scenario of a progressively vanishing perception of pitch when cooling the materal down through the martensitic PT: In the martensite, phase domains and boundaries between twinned domains form an irregular pattern, which can move easily and are good sources of energy dissipation, while the ordered lattice of the more rigid austenite phase responds elastically to external perturbations. More details on the subject of damping in shape memory alloys and their applications can be found, for example, in Refs. \onlinecite{Damping_review,  Damping_NiTiCu, Damping_application} and references therein.

\begin{figure}[ht]
  \centering
  \includegraphics[width=\textwidth]{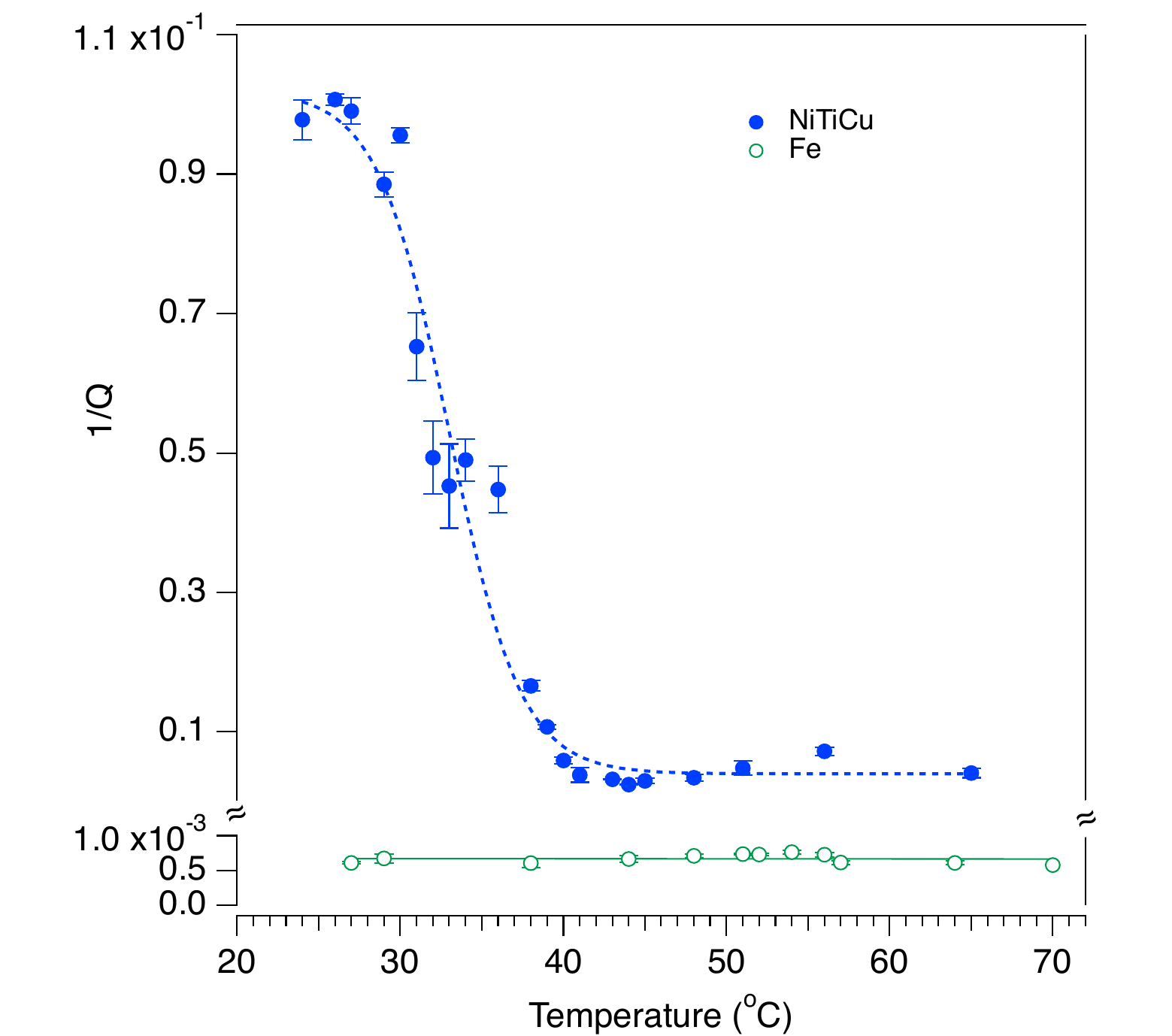}
  \caption{Temperature dependence of the inverse Q factor for the fundamental frequency of nitinol (blue, filled circles) and iron (green, empty circles) rods. The  lines (dashed and solid for nitinol and iron, respectively) are guides for the eyes.}
  \label{fig:Q}
\end{figure}

In order to track the change in the dissipated energy as a function of temperature, we evaluated the $Q$ factor of the fundamental resonance (see the definition in App.~\ref{sec:Appendix2}). Fig.~\ref{fig:Q} shows $Q^{-1}$ (which is a direct indicator of dissipation) as a function of temperature for the NiTiCu and Fe rods. While the value for the iron sample is almost constant, it changes by about an order of magnitude in the NiTiCu sample in the considered temperature range. Although the uncertainty in the estimation of Q is higher than the one for the frequency, especially at low temperatures, our data clearly show that, for decreasing temperatures, Q decreases, which corresponds to the decrease in the decay time.

These data show how the very different damping properties of the martensite and the austenite phases can be easily and pedagogically exposed by studying the decay time of sound recordings. The steps for a quantitative estimation of $Q$ are detailed in App.~\ref{sec:Appendix2}. However, the Audacity software allows the user to simply visualize the wave form in a dB scale, as shown in Fig.~\ref{fig:damp}, which can be a useful tool in a pedagogical context.

\subsection{Remarks about the experimental setup}\label{sec:Experiment}

One should bear in mind that thin bars are not optimized for sound radiation. For this reason, while the sensation of sound and pitch is quite striking when listening to the recording, it diminishes rapidly in the room and becomes barely perceptible at a distance larger than about 1 m from the vibrating bar. Suspending the bars with thin strings attached at the vibration nodes permits the longest ringing times and best approximates the free-boundary conditions that were assumed in the analysis. However, we have also explored other setups and procedures, each having advantages and disadvantages.

For instance, by firmly clamping the bars at one nodal point, one can increase the dominance of the fundamental mode, at the price of stronger damping, i.e. smaller signal-to-noise ratio in an ordinarily quiet room [about 40 dB(A)]. The microphone placement is more critical in this case, compared to freely vibrating bars.

Alternatively, to achieve quicker handling, the bars could be gently held at a nodal point by the experimenter's fingers (suitably gloved for heat insulation) and even hit with the other hand's bare knuckles.\cite{video-nitinol-nanolab} This handling has the pedagogical advantage of adding a sense of touch to the experiment. It also makes it possible to keep the bar close to the experimenter's ear, where its sound is better perceived, and it minimizes waiting time compared to the fixed clamped device. However, the amount of friction at the holding point cannot be precisely controlled and will affect the accuracy in the damping measurement. In all cases, it is suggested to find and mark in bright color the nodal points and the middle point of the bar.

In order to simplify the sound signal processing and analysis, we recommend using bars with rectangular instead of square cross sections. A larger aspect ratio helps keep the frequencies of the bending vibrations in the two dimensions perpendicular to the bar length far apart from each other and prevents the appearance of beats during the decay phase, which may obscure the exponential decay envelope of each mode.

Moreover it is useful to have someone with a musically trained ear assisting, to help assess the quality of the sound while  recording, and spot interference from the environment in the frequency band 300-800~Hz. Finally, be aware that the decay time constant hugely varies both with materials and across temperatures. For example, for our NiTiCu samples, it spans an interval from about 300~ms down to 13~ms, while for iron it is stable at about 600~ms. When recording multiple strikes, be sure to allow the sound to be completely damped before proceeding to the next strike, while monitoring the temperature.

\section{Conclusions}\label{sec:Concl}

We have shown that the martensitic phase transition can be investigated through the acoustic properties of a NiTiCu thin bar. With a minimal setup, students observe that the elasticity of a NiTiCu bar depends critically on its phase. When the transition temperature is crossed, the frequency markedly changes, corresponding to a steep variation in the Young's modulus of the material. By comparing the temperature dependence of the acoustic behavior of NiTiCu and iron, students can realize the importance of the martensitic phase transition to explain the effect. This conveys the idea that the material's macroscopic properties arise at the nano-scale. Moreover, it provides an easy 'real-time' way to measure the elastic properties of a material, which are usually only described in static terms and in an engineering context. It also allows students to access multiple indicators of the changes occurring within the material, both energy conserving (elasticity) and energy dissipating (damping).

Although the detailed microscopic analysis of this phenomenon is out of the scope of the present work, the possibility of capturing a change in the unit cell symmetry just by listening to the sound made when a bar is hit is a great stimulus to discuss the relation between the properties of matter at the atomic level and their macroscopic behavior. The rather simple equipment gives students full operative control, which makes the present experimental approach suited for a Modern Physics Laboratory course both in a physics major and in interdisciplinary curricula.

We finally note that martenistic PTs are usually presented and discussed in the context of the SME. This is indeed a fascinating subject for students, but its complex relationship to the PT is somewhat cumbersome and not so easy to explain. To this respect, the present approach, which does not rely on an explicit discussion of SME, has the advantage of a more straightforward link between the vibro-acoustic properties, with no visible macroscopic change in the material, and the features of the PT itself. 


\section*{Supporting Material}
Please click on this link to access the supplementary material, which includes audio recordings of the sounds described in the main text. Print readers can see the supplementary material at [DOI to be inserted by AIPP].
\begin{enumerate}
\item Audio file {\em S1\_Drop\_sounds.mp4}. Sounds produced by dropping NiTiCu bars in their austenite and martensite phase on the floor.
\item Audio file {\em S2\_Temperature\_sounds.mp4}. Sounds produced by hitting iron and NiTiCu bars when their temperature is lowered from 50$^\circ$C to 30$^\circ$C.
\end{enumerate}

\begin{acknowledgments}
G.G. acknowledges partial support from project Unimore FAR 2024 NT-ROBOT. C.A.R. acknowledges support by the Next Generation EU PRIN 2022 grant n. 202284JP34 (VIBETWO). V.D.R. acknowledges partial support by the Next Generation EU PRIN 2022 grant n. 2022NXLTYN (TUNES). We gratefully thanks Rahel Goldoni for assistance with the artwork.
\end{acknowledgments}

\section*{Author Declarations}
The authors have no conflicts to disclose. All authors contributed equally to this work.

\appendix

\section{Acoustics of thin bars}\label{sec:Appendix}

As a first approximation, the bending normal modes of a free thin bar with uniform cross section are calculated in the Euler-Bernoulli beam approximation, i.e. assuming small deformations, negligible shear, and that the deformation is pure bending. The frequencies of the normal modes can be expressed as the product of three factors accounting respectively for the shape, size and material of the bar. Under these assumptions, the fundamental frequency for the bending vibration of a free bar having a constant cross-section is
\begin{equation}\label{eq:barfreq}
  \nu_0=A\frac{h}{L^2}c,
\end{equation}
where $L$ is the length of the bar, $h$ its thickness in the direction of the strike, and $A$ is a constant accounting for the shape of the cross-section. The only factor depending on the cross section in Eq.~\eqref{eq:barfreq} is $A$. For a bar with rectangular cross section, $A=1.028$. For a circular cross section, $A=0.890$, in which case $h$ represents the bar's diameter. $c$ is the speed of a longitudinal elastic wave traveling in the bulk material, ``speed of sound" in brief, which only depends on two properties of the material itself, namely its density $\rho$ and its Young's modulus $E$ as
\begin{equation}\label{eq:speed}
  c=\sqrt{\frac{E}{\rho}}.
\end{equation}
The higher frequency modes are in non-harmonic ratios to the fundamental frequency. The constant $A$ is different for each overtone, but it depends only on the boundary conditions and shape of the mode, and not on the material properties, so that the overtone frequency ratios do not depend on temperature.

This simple model is sufficiently accurate in the slender beam limit, i.e. when $\lambda \gg 2\pi K$, where $K$ is the radius of gyration of the beam section, and $\lambda$ the wavelength of the mode. This is the case for our test bar, since, at least for the fundamental mode, it yields
\begin{equation}
  \frac{2\pi K}{\lambda}\approx\frac{\pi h}{\sqrt{12}L}=0.003.
\end{equation}
As a reference, the expected fundamental frequency for our iron bar can be found by substituting in Eq.~\eqref{eq:speed} tabulated data for this material ($\rho = 7800$~kg~m$^{-3}$ and $E = 211$~GPa) and the circular section shape factor for $A$. Eq.~\eqref{eq:barfreq} then returns a frequency of $895$\,Hz, within about $2\%$ from the measured one of $876$\,Hz. The iron sample at room temperature produces a clearly ringing A$_5$ note (which is nominally at 880 Hz).

We can make the temperature dependence of $\nu_0$ explicit by substituting in Eq.~\eqref{eq:speed} the density. Therefore, assuming the shape of the bar remains unchanged, the fundamental frequency, within the Euler-Bernoulli model, can be expressed as a function of temperature-dependent variables as
\begin{equation}
  \nu(T)\propto h^2(T)\sqrt{\frac{E(T)}{L^3(T)}}.
\end{equation}

The position of the nodal points of each vibration mode can also be analytically determined for the ideal thin bar model. For a given bar geometry, the shapes of the vibration modes (and thereby the positions of the nodes) only depend on the boundary conditions, while the modal frequency also depends on the material's properties. The nodes of the fundamental mode are located at a distance of $0.224\cdot L$ from each end of the bar~\cite{BarModes} no matter what the fundamental frequency is. This fact can be exploited in the experiments to enhance the pitch sensation. Indeed, the sound spectrum of the struck bar is composed of a wide-band impact noise, and of a sequence of distinct overtones, each corresponding to a single free mode of vibration of the bar. Since the overtones are not in harmonic ratios with the fundamental, they may disrupt or substantially alter the overall sensation of pitch. However by holding the bar at a nodal point, all modes that do not have a node at that point are quickly damped, and the fundamental stands out more clearly.

\section{Estimation of the Q factor}\label{sec:Appendix2}

The recorded impact sound must be analyzed to find the peak corresponding to the bar's first vibration mode. The raw sound will most probably contain background noise. In order to obtain a simple waveform closer to the one for a damped harmonic oscillator, it is useful to apply a narrow band-pass filter using, for instance, Audacity's built-in Filter-curve effect, which includes an intuitive interface to design the filter. We applied a 1/3 octave band filter of order 3, which was sufficient to get rid of the background noise and isolate the first bending mode of vibration perpendicular to the direction of the strike.

Internal damping results in an exponential decay in the sound oscillation amplitude. It can be modeled by adding an effective viscous term to the equation of motion for the free vibration
\begin{equation}
  \frac{d^2 x(t)}{dt^2} + 2\zeta\omega_0\frac{dx(t)}{dt} + \omega_0^2 x(t)=0,
\end{equation}
with solution
\begin{equation}
  x(t) = x_0\cos[\omega (t - t_0)]\exp\left[- \frac{(t - t_0)}{\tau}\right],
\end{equation}
where $x(t)$ is the sound amplitude, $t_0$ is the time at which the amplitude reaches its maximum value $x_0$, $\omega_0$ is the undamped fundamental angular frequency, $\zeta$ the damping ratio, $\omega = \omega_0\sqrt{1-\zeta^2}$ and $\tau = \frac{1}{\zeta\omega_0}$ the time constant for the exponential decay.

The damping ratio $\zeta$ can be calculated analytically from the logarithmic decrement $\delta$ of the signal
\begin{equation}
  \delta = \frac{1}{n}\log\left(\frac{x_0}{x_n}\right),
\end{equation}
where $x_n$ is the amplitude of the signal after $n$ full periods of oscillation from the reference time $t_0$.
If the signal's amplitude is given in dB, as it is customary, $\delta$ can be obtained from a simple linear fit of the envelope 
\begin{equation}
  \delta = \frac{0.1151}{n}(A_0-A_n),
\end{equation}
where $A_i = 20\log_{10}(x_i)$ are the signal amplitudes in dB. Then
\begin{equation}
  \zeta = \frac{\delta}{\sqrt{4\pi^2+\delta^2}} \approx \frac{\delta}{2\pi}.
\end{equation}

Finally, damping is often quantified with a quality factor $Q$, which is simply
\begin{equation}
  Q = \frac{1}{2\zeta} \approx \frac{\pi}{\delta}.
\end{equation}

\bibliography{biblio_nitinol}

\begin{thebibliography}{36}
\expandafter\ifx\csname natexlab\endcsname\relax\def\natexlab#1{#1}\fi
\expandafter\ifx\csname bibnamefont\endcsname\relax
  \def\bibnamefont#1{#1}\fi
\expandafter\ifx\csname bibfnamefont\endcsname\relax
  \def\bibfnamefont#1{#1}\fi
\expandafter\ifx\csname citenamefont\endcsname\relax
  \def\citenamefont#1{#1}\fi
\expandafter\ifx\csname url\endcsname\relax
  \def\url#1{\texttt{#1}}\fi
\expandafter\ifx\csname urlprefix\endcsname\relax\def\urlprefix{URL }\fi
\providecommand{\bibinfo}[2]{#2}
\providecommand{\eprint}[2][]{\url{#2}}

\bibitem[{\citenamefont{Lisotti et~al.}(2013)\citenamefont{Lisotti, Renzi,
  Rozzi, Villa, Albertini, and Goldoni}}]{nanolab}
\bibinfo{author}{\bibfnamefont{A.}~\bibnamefont{Lisotti}},
  \bibinfo{author}{\bibfnamefont{V.~D.} \bibnamefont{Renzi}},
  \bibinfo{author}{\bibfnamefont{C.~A.} \bibnamefont{Rozzi}},
  \bibinfo{author}{\bibfnamefont{E.}~\bibnamefont{Villa}},
  \bibinfo{author}{\bibfnamefont{F.}~\bibnamefont{Albertini}},
  \bibnamefont{and} \bibinfo{author}{\bibfnamefont{G.}~\bibnamefont{Goldoni}},
  \bibinfo{journal}{Physics Education} \textbf{\bibinfo{volume}{48}},
  \bibinfo{pages}{298} (\bibinfo{year}{2013}),
  \urlprefix\url{https://dx.doi.org/10.1088/0031-9120/48/3/298}.

\bibitem[{\citenamefont{Otsuka and Ren}(2005)}]{NiTiCrystal}
\bibinfo{author}{\bibfnamefont{K.}~\bibnamefont{Otsuka}} \bibnamefont{and}
  \bibinfo{author}{\bibfnamefont{X.}~\bibnamefont{Ren}},
  \bibinfo{journal}{Progress in Materials Science}
  \textbf{\bibinfo{volume}{50}}, \bibinfo{pages}{511} (\bibinfo{year}{2005}),
  ISSN \bibinfo{issn}{0079-6425},
  \urlprefix\url{http://www.sciencedirect.com/science/article/pii/S0079642504000647}.

\bibitem[{\citenamefont{{Clapp, P. C.}}(1995)}]{Def2}
\bibinfo{author}{\bibnamefont{{Clapp, P. C.}}}, \bibinfo{journal}{J. Phys. IV
  France} \textbf{\bibinfo{volume}{05}}, \bibinfo{pages}{C8}
  (\bibinfo{year}{1995}), \urlprefix\url{https://doi.org/10.1051/jp4:1995802}.

\bibitem[{vid({\natexlab{a}})}]{video-SME}
\emph{\bibinfo{title}{Shape memory effect at nanolab.unimore.it}},
  \bibinfo{note}{accessed 2023-09-14},
  \urlprefix\url{https://www.youtube.com/watch?v=9DH-VOILyWE}.

\bibitem[{\citenamefont{Amariei et~al.}(2010)\citenamefont{Amariei,
  Frunzaverde, Vela, and Gillich}}]{Edu2}
\bibinfo{author}{\bibfnamefont{D.}~\bibnamefont{Amariei}},
  \bibinfo{author}{\bibfnamefont{D.}~\bibnamefont{Frunzaverde}},
  \bibinfo{author}{\bibfnamefont{I.}~\bibnamefont{Vela}}, \bibnamefont{and}
  \bibinfo{author}{\bibfnamefont{G.~R.} \bibnamefont{Gillich}},
  \bibinfo{journal}{Procedia - Social and Behavioral Sciences}
  \textbf{\bibinfo{volume}{2}}, \bibinfo{pages}{5104} (\bibinfo{year}{2010}),
  ISSN \bibinfo{issn}{1877-0428}, \bibinfo{note}{innovation and Creativity in
  Education},
  \urlprefix\url{https://www.sciencedirect.com/science/article/pii/S1877042810008694}.

\bibitem[{\citenamefont{Song and Bannerot}(2007)}]{Edu3}
\bibinfo{author}{\bibfnamefont{G.}~\bibnamefont{Song}} \bibnamefont{and}
  \bibinfo{author}{\bibfnamefont{R.}~\bibnamefont{Bannerot}}, in
  \emph{\bibinfo{booktitle}{2007 Annual Conference \& Exposition}}
  (\bibinfo{publisher}{ASEE Conferences}, \bibinfo{address}{Honolulu, Hawaii},
  \bibinfo{year}{2007}),
  \urlprefix\url{https://peer.asee.org/development-of-an-interactive-shape-memory-alloy-demonstration-for-smart-materials-curricula}.

\bibitem[{\citenamefont{Chaudhari et~al.}(2021)\citenamefont{Chaudhari, Vora,
  and Parikh}}]{ReviewNiTiApplications}
\bibinfo{author}{\bibfnamefont{R.}~\bibnamefont{Chaudhari}},
  \bibinfo{author}{\bibfnamefont{J.~J.} \bibnamefont{Vora}}, \bibnamefont{and}
  \bibinfo{author}{\bibfnamefont{D.~M.} \bibnamefont{Parikh}}, in
  \emph{\bibinfo{booktitle}{Recent Advances in Mechanical Infrastructure}},
  edited by \bibinfo{editor}{\bibfnamefont{A.~K.} \bibnamefont{Parwani}},
  \bibinfo{editor}{\bibfnamefont{P.}~\bibnamefont{Ramkumar}},
  \bibinfo{editor}{\bibfnamefont{K.}~\bibnamefont{Abhishek}}, \bibnamefont{and}
  \bibinfo{editor}{\bibfnamefont{S.~K.} \bibnamefont{Yadav}}
  (\bibinfo{publisher}{Springer Singapore}, \bibinfo{address}{Singapore},
  \bibinfo{year}{2021}), pp. \bibinfo{pages}{123--132}, ISBN
  \bibinfo{isbn}{978-981-33-4176-0},
  \urlprefix\url{https://link.springer.com/chapter/10.1007/978-981-33-4176-0_10}.

\bibitem[{rec()}]{recordings}
\bibinfo{note}{Sound recordings are available as Supplementary Material, file
  {\em S2\_Temperature\_sounds.mp4}}.

\bibitem[{sto()}]{storica}
\bibinfo{note}{This fact was quickly noticed in early experiments by
  \citet{NOLReport}, who made good use of their ears to characterize the
  samples and supplement their measurements. Curiously, they classified the
  sounds they perceived using vivid, yet imprecise terms like ``dull", ``slight
  ring", ``clear ring", ``sharp high frequency ring" and ``transition".
  However, they did not pursue their analysis beyond this.}

\bibitem[{\citenamefont{Gisser et~al.}(1994)\citenamefont{Gisser, Geselbracht,
  Cappellari, Hunsberger, Ellis, Perepezko, and Lisensky}}]{NiTiTemp}
\bibinfo{author}{\bibfnamefont{K.~R.~C.} \bibnamefont{Gisser}},
  \bibinfo{author}{\bibfnamefont{M.~J.} \bibnamefont{Geselbracht}},
  \bibinfo{author}{\bibfnamefont{A.}~\bibnamefont{Cappellari}},
  \bibinfo{author}{\bibfnamefont{L.}~\bibnamefont{Hunsberger}},
  \bibinfo{author}{\bibfnamefont{A.~B.} \bibnamefont{Ellis}},
  \bibinfo{author}{\bibfnamefont{J.}~\bibnamefont{Perepezko}},
  \bibnamefont{and} \bibinfo{author}{\bibfnamefont{G.~C.}
  \bibnamefont{Lisensky}}, \bibinfo{journal}{Journal of Chemical Education}
  \textbf{\bibinfo{volume}{71}}, \bibinfo{pages}{334 } (\bibinfo{year}{1994}),
  \urlprefix\url{http://dx.doi.org/10.1021/ed071p334}.

\bibitem[{\citenamefont{Campbell et~al.}(2014)\citenamefont{Campbell, Peterson,
  and Fitzjarrald}}]{spectroscopy}
\bibinfo{author}{\bibfnamefont{D.~J.} \bibnamefont{Campbell}},
  \bibinfo{author}{\bibfnamefont{J.~P.} \bibnamefont{Peterson}},
  \bibnamefont{and} \bibinfo{author}{\bibfnamefont{T.~J.}
  \bibnamefont{Fitzjarrald}}, \bibinfo{journal}{Journal of Chemical Education}
  \textbf{\bibinfo{volume}{91}}, \bibinfo{pages}{1684} (\bibinfo{year}{2014}),
  ISSN \bibinfo{issn}{0021-9584},
  \urlprefix\url{https://doi.org/10.1021/ed500070j}.

\bibitem[{Not({\natexlab{a}})}]{Note3}
\bibinfo{note}{Though, in phase transitions, we typically associate the
  low-temperature phase with more symmetric structures, this is not necessarily
  always the case. In fact, in our system, the high-temperature B2 phase has a
  formation enthalpy which is about $0.034$~eV/atom larger than the
  low-temperature B19$^\prime$ phase (see Ref. \onlinecite{Ko2015}). The B19'
  phase is therefore stabilized at sufficiently low temperatures.}

\bibitem[{\citenamefont{{Van Humbeeck}}(2003)}]{Damping_capacity}
\bibinfo{author}{\bibfnamefont{J.}~\bibnamefont{{Van Humbeeck}}},
  \bibinfo{journal}{Journal of Alloys and Compounds}
  \textbf{\bibinfo{volume}{355}}, \bibinfo{pages}{58} (\bibinfo{year}{2003}),
  ISSN \bibinfo{issn}{0925-8388}, \bibinfo{note}{proceedings of the
  International Symposium on High Damping Materials},
  \urlprefix\url{https://www.sciencedirect.com/science/article/pii/S0925838803002688}.

\bibitem[{\citenamefont{Chang and Read}(1951)}]{FirstSME}
\bibinfo{author}{\bibfnamefont{L.~C.} \bibnamefont{Chang}} \bibnamefont{and}
  \bibinfo{author}{\bibfnamefont{T.~A.} \bibnamefont{Read}},
  \bibinfo{journal}{Transactions of the American Institute of Mining
  Metallurgical and Petroleum Engineers} \textbf{\bibinfo{volume}{191}}
  (\bibinfo{year}{1951}),
  \urlprefix\url{https://archive.org/details/american-institute-of-mining-and-metallurgical_1951_191/page/46/mode/2up}.

\bibitem[{\citenamefont{Hodgson and Biermann}(1990)}]{SMEAlloys}
\bibinfo{author}{\bibfnamefont{D.~E.} \bibnamefont{Hodgson}} \bibnamefont{and}
  \bibinfo{author}{\bibfnamefont{R.~J.} \bibnamefont{Biermann}},
  \emph{\bibinfo{title}{ASM Handbook: Nonferrous Alloys and Special-Purpose
  Materials}} (\bibinfo{publisher}{ASM International}, \bibinfo{year}{1990}),
  vol.~\bibinfo{volume}{2}, p. \bibinfo{pages}{2524}, ISBN
  \bibinfo{isbn}{978-0-87170-378-1},
  \urlprefix\url{https://dl.asminternational.org/handbooks/edited-volume/14/chapter-abstract/188049/Shape-Memory-Alloys}.

\bibitem[{\citenamefont{Buehler et~al.}(1963)\citenamefont{Buehler, Gilfrich,
  and Wiley}}]{MostSME}
\bibinfo{author}{\bibfnamefont{W.~J.} \bibnamefont{Buehler}},
  \bibinfo{author}{\bibfnamefont{J.~V.} \bibnamefont{Gilfrich}},
  \bibnamefont{and} \bibinfo{author}{\bibfnamefont{R.~C.} \bibnamefont{Wiley}},
  \bibinfo{journal}{Journal of Applied Physics} \textbf{\bibinfo{volume}{34}},
  \bibinfo{pages}{1475} (\bibinfo{year}{1963}),
  \urlprefix\url{http://dx.doi.org/10.1063/1.1729603}.

\bibitem[{\citenamefont{Ko et~al.}(2015)\citenamefont{Ko, Grabowski, and
  Neugebauer}}]{Ko2015}
\bibinfo{author}{\bibfnamefont{W.~S.} \bibnamefont{Ko}},
  \bibinfo{author}{\bibfnamefont{B.}~\bibnamefont{Grabowski}},
  \bibnamefont{and}
  \bibinfo{author}{\bibfnamefont{J.}~\bibnamefont{Neugebauer}},
  \bibinfo{journal}{Phys. Rev. B - Condens. Matter Mater. Phys.}
  \textbf{\bibinfo{volume}{92}}, \bibinfo{pages}{1} (\bibinfo{year}{2015}),
  ISSN \bibinfo{issn}{1550235X},
  \urlprefix\url{https://journals.aps.org/prb/abstract/10.1103/PhysRevB.92.134107}.

\bibitem[{\citenamefont{Otsuka and Wayman}(1999)}]{SMABook}
\bibinfo{editor}{\bibfnamefont{K.}~\bibnamefont{Otsuka}} \bibnamefont{and}
  \bibinfo{editor}{\bibfnamefont{C.~M.} \bibnamefont{Wayman}}, eds.,
  \emph{\bibinfo{title}{Shape memory materials}} (\bibinfo{publisher}{Cambridge
  University Press}, \bibinfo{year}{1999}), ISBN \bibinfo{isbn}{9780521663847},
  \urlprefix\url{http://www.cambridge.org/it/academic/subjects/engineering/materials-science/shape-memory-materials#gVjol5kZADu6G7fF.97}.

\bibitem[{Pha(2007)}]{PhaseTransfChap4}
in \emph{\bibinfo{booktitle}{Phase Transformations}}, edited by
  \bibinfo{editor}{\bibfnamefont{S.}~\bibnamefont{Banerjee}} \bibnamefont{and}
  \bibinfo{editor}{\bibfnamefont{P.}~\bibnamefont{Mukhopadhyay}}
  (\bibinfo{publisher}{Pergamon}, \bibinfo{year}{2007}),
  vol.~\bibinfo{volume}{12} of \emph{\bibinfo{series}{Pergamon Materials
  Series}}, pp. \bibinfo{pages}{257--376},
  \urlprefix\url{https://www.sciencedirect.com/science/article/pii/S1470180407800575}.

\bibitem[{\citenamefont{Nespoli et~al.}(2011)\citenamefont{Nespoli, Villa, and
  Besseghini}}]{IENIbars}
\bibinfo{author}{\bibfnamefont{A.}~\bibnamefont{Nespoli}},
  \bibinfo{author}{\bibfnamefont{E.}~\bibnamefont{Villa}}, \bibnamefont{and}
  \bibinfo{author}{\bibfnamefont{S.}~\bibnamefont{Besseghini}},
  \bibinfo{journal}{Journal of Alloys and Compounds}
  \textbf{\bibinfo{volume}{509}}, \bibinfo{pages}{644} (\bibinfo{year}{2011}),
  ISSN \bibinfo{issn}{0925-8388},
  \urlprefix\url{https://www.sciencedirect.com/science/article/pii/S092583881002373X}.

\bibitem[{Aud()}]{Audacity}
\emph{\bibinfo{title}{{Audacity(R)}}}, \bibinfo{note}{accessed 2023-09-14},
  \urlprefix\url{http://audacity.sourceforge.net}.

\bibitem[{\citenamefont{Nix and MacNair}(1941)}]{LinThermExp}
\bibinfo{author}{\bibfnamefont{F.~C.} \bibnamefont{Nix}} \bibnamefont{and}
  \bibinfo{author}{\bibfnamefont{D.}~\bibnamefont{MacNair}},
  \bibinfo{journal}{Phys. Rev.} \textbf{\bibinfo{volume}{60}},
  \bibinfo{pages}{597} (\bibinfo{year}{1941}),
  \urlprefix\url{https://link.aps.org/doi/10.1103/PhysRev.60.597}.

\bibitem[{\citenamefont{Ledbetter and Reed}(1973)}]{YoungTemp}
\bibinfo{author}{\bibfnamefont{H.~M.} \bibnamefont{Ledbetter}}
  \bibnamefont{and} \bibinfo{author}{\bibfnamefont{R.~P.} \bibnamefont{Reed}},
  \bibinfo{journal}{Journal of Physical and Chemical Reference Data}
  \textbf{\bibinfo{volume}{2}}, \bibinfo{pages}{531} (\bibinfo{year}{1973}),
  ISSN \bibinfo{issn}{0047-2689},
  \eprint{https://pubs.aip.org/aip/jpr/article-pdf/2/3/531/19257316/531\_1\_online.pdf},
  \urlprefix\url{https://doi.org/10.1063/1.3253127}.

\bibitem[{nor()}]{normalExpansion}
\bibinfo{note}{All metals and the vast majority of metallic alloys have a
  positive thermal expansion coefficient.}

\bibitem[{Not({\natexlab{b}})}]{NoteUnitCell}
\bibinfo{note}{From the standard unit cell parameters in the tetragonal and
  monoclinic phases of NiTi reported by \citet{NiTiCrystal}, one obtains a
  theoretical density of 6.46 g/cm$^3$ for the austenite, and 6.17 g/cm$^3$ for
  the martensite. This results in an effective thermal coefficient of the order
  of 10$^{-5}$ $^\circ$C$^{-1}$.}

\bibitem[{\citenamefont{Mazzolai et~al.}(2003)\citenamefont{Mazzolai,
  Biscarini, Campanella, Coluzzi, Mazzolai, Rotini, and Tuissi}}]{E_niti}
\bibinfo{author}{\bibfnamefont{F.}~\bibnamefont{Mazzolai}},
  \bibinfo{author}{\bibfnamefont{A.}~\bibnamefont{Biscarini}},
  \bibinfo{author}{\bibfnamefont{R.}~\bibnamefont{Campanella}},
  \bibinfo{author}{\bibfnamefont{B.}~\bibnamefont{Coluzzi}},
  \bibinfo{author}{\bibfnamefont{G.}~\bibnamefont{Mazzolai}},
  \bibinfo{author}{\bibfnamefont{A.}~\bibnamefont{Rotini}}, \bibnamefont{and}
  \bibinfo{author}{\bibfnamefont{A.}~\bibnamefont{Tuissi}},
  \bibinfo{journal}{Acta Materialia} \textbf{\bibinfo{volume}{51}},
  \bibinfo{pages}{573} (\bibinfo{year}{2003}), ISSN \bibinfo{issn}{1359-6454},
  \urlprefix\url{https://www.sciencedirect.com/science/article/pii/S1359645402004391}.

\bibitem[{\citenamefont{Rotini et~al.}(2001)\citenamefont{Rotini, Biscarini,
  Campanella, Coluzzi, Mazzolai, and Mazzolai}}]{Ref_E}
\bibinfo{author}{\bibfnamefont{A.}~\bibnamefont{Rotini}},
  \bibinfo{author}{\bibfnamefont{A.}~\bibnamefont{Biscarini}},
  \bibinfo{author}{\bibfnamefont{R.}~\bibnamefont{Campanella}},
  \bibinfo{author}{\bibfnamefont{B.}~\bibnamefont{Coluzzi}},
  \bibinfo{author}{\bibfnamefont{G.}~\bibnamefont{Mazzolai}}, \bibnamefont{and}
  \bibinfo{author}{\bibfnamefont{F.}~\bibnamefont{Mazzolai}},
  \bibinfo{journal}{Scripta Materialia} \textbf{\bibinfo{volume}{44}},
  \bibinfo{pages}{719} (\bibinfo{year}{2001}), ISSN \bibinfo{issn}{1359-6462},
  \urlprefix\url{https://www.sciencedirect.com/science/article/pii/S1359646200006722}.

\bibitem[{\citenamefont{Zotov et~al.}(2014)\citenamefont{Zotov, Marzynkevitsch,
  and Mittemeijer}}]{Zotov2014}
\bibinfo{author}{\bibfnamefont{N.}~\bibnamefont{Zotov}},
  \bibinfo{author}{\bibfnamefont{V.}~\bibnamefont{Marzynkevitsch}},
  \bibnamefont{and} \bibinfo{author}{\bibfnamefont{E.~J.}
  \bibnamefont{Mittemeijer}}, \bibinfo{journal}{Journal of Alloys and
  Compounds} \textbf{\bibinfo{volume}{616}}, \bibinfo{pages}{385}
  (\bibinfo{year}{2014}), ISSN \bibinfo{issn}{0925-8388},
  \urlprefix\url{https://www.sciencedirect.com/science/article/pii/S0925838814017411}.

\bibitem[{\citenamefont{Wagner and Windl}(2008)}]{YoungPhases}
\bibinfo{author}{\bibfnamefont{M.-X.} \bibnamefont{Wagner}} \bibnamefont{and}
  \bibinfo{author}{\bibfnamefont{W.}~\bibnamefont{Windl}},
  \bibinfo{journal}{Acta Materialia} \textbf{\bibinfo{volume}{56}},
  \bibinfo{pages}{6232} (\bibinfo{year}{2008}), ISSN \bibinfo{issn}{1359-6454},
  \urlprefix\url{https://www.sciencedirect.com/science/article/pii/S1359645408006083}.

\bibitem[{Not({\natexlab{c}})}]{NotePorosity}
\bibinfo{note}{In fact, the martensite phase has larger porosity than the
  austenite one, and its surface appears to the eye slightly rougher and less
  polished. However the corresponding change in friction with air is expected
  to be negligible in this context.}

\bibitem[{\citenamefont{Saedi et~al.}(2023)\citenamefont{Saedi, Acar, Raji,
  Saghaian, and Mirsayar}}]{Damping_review}
\bibinfo{author}{\bibfnamefont{S.}~\bibnamefont{Saedi}},
  \bibinfo{author}{\bibfnamefont{E.}~\bibnamefont{Acar}},
  \bibinfo{author}{\bibfnamefont{H.}~\bibnamefont{Raji}},
  \bibinfo{author}{\bibfnamefont{S.~E.} \bibnamefont{Saghaian}},
  \bibnamefont{and} \bibinfo{author}{\bibfnamefont{M.}~\bibnamefont{Mirsayar}},
  \bibinfo{journal}{Journal of Alloys and Compounds}
  \textbf{\bibinfo{volume}{956}}, \bibinfo{pages}{170286}
  (\bibinfo{year}{2023}), ISSN \bibinfo{issn}{0925-8388},
  \urlprefix\url{https://www.sciencedirect.com/science/article/pii/S092583882301589X}.

\bibitem[{\citenamefont{Yoshida et~al.}(2003)\citenamefont{Yoshida, Monma,
  Iino, Otsuka, Asai, and Tsuzuki}}]{Damping_NiTiCu}
\bibinfo{author}{\bibfnamefont{I.}~\bibnamefont{Yoshida}},
  \bibinfo{author}{\bibfnamefont{D.}~\bibnamefont{Monma}},
  \bibinfo{author}{\bibfnamefont{K.}~\bibnamefont{Iino}},
  \bibinfo{author}{\bibfnamefont{K.}~\bibnamefont{Otsuka}},
  \bibinfo{author}{\bibfnamefont{M.}~\bibnamefont{Asai}}, \bibnamefont{and}
  \bibinfo{author}{\bibfnamefont{H.}~\bibnamefont{Tsuzuki}},
  \bibinfo{journal}{Journal of Alloys and Compounds}
  \textbf{\bibinfo{volume}{355}}, \bibinfo{pages}{79} (\bibinfo{year}{2003}),
  ISSN \bibinfo{issn}{0925-8388}, \bibinfo{note}{proceedings of the
  International Symposium on High Damping Materials},
  \urlprefix\url{https://www.sciencedirect.com/science/article/pii/S0925838803002809}.

\bibitem[{\citenamefont{Humbeeck and Kustov}(2005)}]{Damping_application}
\bibinfo{author}{\bibfnamefont{J.~V.} \bibnamefont{Humbeeck}} \bibnamefont{and}
  \bibinfo{author}{\bibfnamefont{S.}~\bibnamefont{Kustov}},
  \bibinfo{journal}{Smart Materials and Structures}
  \textbf{\bibinfo{volume}{14}}, \bibinfo{pages}{S171} (\bibinfo{year}{2005}),
  \urlprefix\url{https://dx.doi.org/10.1088/0964-1726/14/5/001}.

\bibitem[{vid({\natexlab{b}})}]{video-nitinol-nanolab}
\emph{\bibinfo{title}{Nanolab}}, \bibinfo{note}{video description of the
  procedure (in Italian), Accessed 2023-09-14},
  \urlprefix\url{https://www.nanolab.unimore.it/laboratori/lab-nitinolo/transizioni-di-fase-2/transizioni-di-fase-1-resistivita-e-allungamento/}.

\bibitem[{\citenamefont{Lord~Rayleigh}(1877)}]{BarModes}
\bibinfo{author}{\bibfnamefont{J.~W.~S.} \bibnamefont{Lord~Rayleigh}},
  \emph{\bibinfo{title}{The Theory of Sound}} (\bibinfo{publisher}{MacMillan
  {\&} co.}, \bibinfo{year}{1877}), \bibinfo{note}{reprint Dover Publications,
  New York, (1945)}.

\bibitem[{\citenamefont{Buehler and Wiley}(1961)}]{NOLReport}
\bibinfo{author}{\bibfnamefont{W.~J.} \bibnamefont{Buehler}} \bibnamefont{and}
  \bibinfo{author}{\bibfnamefont{R.~C.} \bibnamefont{Wiley}},
  \bibinfo{type}{Tech. Rep.} \bibinfo{number}{AD0266607},
  \bibinfo{institution}{U.S. Naval Ordnance Laboratory}, \bibinfo{address}{Wite
  Oak MD} (\bibinfo{year}{1961}),
  \urlprefix\url{https://apps.dtic.mil/sti/citations/tr/AD0266607}.

\end{thebibliography}
\bibliographystyle{apsrev}


\newpage

\newpage

\newpage

\newpage

\newpage

\newpage

\end{document}